\documentclass[%
twocolumn,
amsmath,amssymb,
aps,
superscriptaddress,
]{revtex4-2}

\usepackage{url}
\usepackage{graphicx}
\graphicspath{{.}{figs/}{../figs/}}
\usepackage{afterpage}
\usepackage{siunitx}
\DeclareSIUnit\angstrom{\text {Å}}
\usepackage{dcolumn}
\usepackage{bm}
\usepackage{tikz}
\usetikzlibrary{positioning,intersections,angles,quotes}
\usepackage[caption=false]{subfig}
\usepackage{booktabs} 
\usepackage{comment}
\usepackage{physics}
\usepackage{mathtools}
\usepackage{enumerate}

\usepackage[colorlinks=true,citecolor=blue,linkcolor=magenta]{hyperref}

\usepackage{longtable}
\usepackage{booktabs}

\begin{document}

\title{Improving threshold for fault-tolerant color code quantum computing by flagged weight optimization}

\author{Yugo Takada}
\email{u751105k@ecs.osaka-u.ac.jp}
\affiliation{%
  Graduate School of Engineering Science, Osaka University, 1-3 Machikaneyama, Toyonaka, Osaka 560-8531, Japan
}%
\author{Keisuke Fujii}%
\email{fujii@qc.ee.es.osaka-u.ac.jp}
\affiliation{%
  Graduate School of Engineering Science, Osaka University, 1-3 Machikaneyama, Toyonaka, Osaka 560-8531, Japan
}%
\affiliation{%
  Center for Quantum Information and Quantum Biology,
  Osaka University, 1-2 Machikaneyama, Toyonaka 560-0043, Japan
}%
\affiliation{%
  RIKEN Center for Quantum Computing (RQC),
  Hirosawa 2-1, Wako, Saitama 351-0198, Japan
}%
%
\date{\today}
\begin{abstract}
  Color codes are promising quantum error correction (QEC) codes because they have an advantage over surface codes in that all Clifford gates can be implemented transversally.
  However, thresholds of color codes under circuit-level noise are relatively low mainly because measurements of their high-weight stabilizer generators cause an increase in a circuit depth, and thus, substantial errors are introduced. This makes color codes not the best candidate for fault-tolerant quantum computing.
  Here, we propose a method to suppress the impact of such errors by optimizing weights of decoders using conditional error probabilities conditioned on the measurement outcomes of flag qubits.
  In numerical simulations, we improve the threshold of the (4.8.8) color code under circuit-level noise from 0.14\% to around 0.27\%, which is calculated by using an integer programming decoder.
  Furthermore, in the (6.6.6) color code, we achieve a circuit-level threshold of around 0.36\%, which is almost the same value as the highest value in the previous studies employing the same noise model.
  In both cases, an effective code distance is also improved compared to a conventional method that uses a single ancilla qubit for each stabilizer measurement. Thereby, the achieved logical error rates at low physical error rates are almost one order of magnitude lower than those of the conventional method with the same code distance. Even when compared to the single ancilla method with higher code distance, considering the increased number of qubits used in our method, we achieve lower logical error rates in most cases.
  This method can also be applied to other weight-based decoders, making the color codes more promising for the candidate of experimental implementation of QEC.
  Furthermore, one can utilize this approach to improve a threshold of wider classes of QEC codes, such as high-rate quantum low-density parity check codes.
\end{abstract}

\maketitle

\section{Introduction}
Quantum computers have the potential to efficiently solve computationally difficult problems, such as factorization of large numbers~\cite{shor} and simulations of quantum many-body systems~\cite{Feynman1982}.
However, qubits are highly susceptible to noise, making it difficult to perform accurate quantum computations.
Quantum error correction (QEC) is a critical solution to suppress the impact of noise, enabling fault-tolerant quantum computing (FTQC) by encoding fragile physical qubits into robust logical qubits through quantum error correction codes (QECCs).
In the theory of QEC, if an error probability per quantum gate is below a certain threshold, we can perform arbitrarily accurate quantum computations by increasing the number of physical qubits.
Consequently, extensive research has been undertaken to establish FTQC protocols with a high threshold.

Currently, surface codes \cite{surface} are considered to be one of the most promising QECCs, as they have been experimentally demonstrated in recent years \cite{krinner2022realizing,zhao2022realization,acharya2022suppressing,bluvstein2023logical}.
The notable advantages of surface codes are their ease of physical implementation as well as their high thresholds.
The thresholds of surface codes under circuit-level noise are estimated to be around 0.5\%-1.1\% \cite{surface_circuit_threshold1,surface_circuit_threshold2,chamberland2020triangular,surface_circuit_threshold3,surface_circuit_threshold4,surface_circuit_threshold5}, depending on each noise assumption and way of performing QEC.
On the other hand, surface codes also have a drawback in terms of fault-tolerant implementation of logical gates.
In order to realize large-scale FTQC, it is needed to implement a universal set of logical gates fault-tolerantly with low spatial and temporal overheads.
However, even for certain Clifford gates, fault-tolerantly implementing logical gates using surface codes requires costly techniques \cite{lattice_surgery1,lattice_surgery2,lattice_surgery3}, which can lead to significant overheads.

Another promising QECC is color codes \cite{color1,color2,kesselring2018boundaries}, which admit transversal implementation of all logical Clifford gates due to their high symmetry of stabilizer operators \cite{PhysRevA.91.032330}.
This property has led to color codes being considered as promising QECCs for achieving FTQC with low overhead.
Experimentally, color codes with small code distances have been demonstrated in recent years \cite{PhysRevX.11.041058,ryan2022implementing,bluvstein2023logical}.
However, low thresholds of color codes have made the practical implementation of color code-based FTQC difficult.
For two typical color codes, the (4.8.8) color code and the (6.6.6) color code, the thresholds under circuit-level noise are around 0.08\%-0.14\% \cite{landahl,graphmatching} and 0.2\%-0.47\% \cite{chamberland2020triangular,baireuther2019neural,PRXQuantum.2.020341,lee2024color,zhang2023facilitating}, respectively.
In the [[4,2,2]]-concatenated toric code, which is a subsystem version of the (4.8.8) color code, a circuit-level threshold of 0.41\% has been obtained \cite{criger2016noise} due to the utilization of the gauge degree of freedom, but it does not support transversal $H$ and $S$ gate.
The main cause of the low thresholds in color codes is that stabilizer generators of color codes are high-weight; in other words, they act on many data qubits.
High-weight stabilizer generators cause an increase in the circuit depth of a syndrome measurement circuit, and thus, substantial errors are introduced.

A threshold is also influenced by the performance of decoders.
In particular, for weight-based decoders such as the minimum-weight perfect matching (MWPM) decoder \cite{planar}, the weighted-union find decoder \cite{weighted_union_find}, and the integer programming decoder \cite{landahl}, the optimality of weights has a significant effect on the threshold.
In the conventional way of setting a weight \cite{planar}, the weight $w$ is defined as $w=-\log (p/(1-p))$, where $p$ is a probability of each data error or measurement error.
This weight is not optimal in the sense that it fails to account for the impact of correlated errors, such as hook errors \cite{planar}.
This is because, as detailed in appendix \ref{appendix:a}, this way of setting weight is derived under the independence assumption for each data error or measurement error.
In Ref.~\cite{chamberland2020triangular}, a method is proposed to account for the impact of correlated errors in color codes by adding additional edges in the decoding graph.
However, it is not obvious how to apply the method to color codes other than the (6.6.6) color code or to weight-based decoders other than the Restriction Decoder \cite{unionfind}.

In this paper, we propose {\it flagged weight optimization} (FWO), a method to improve thresholds of color codes under circuit-level noise by optimizing the weights of a decoder using conditional error probabilities conditioned on the measurement outcomes of flag qubits. 
A flag qubit \cite{Yoder2017surfacecodetwist,PhysRevLett.121.050502,Chamberland2018flagfaulttolerant,PhysRevA.101.012342,PRXQuantum.1.010302} is an additional ancilla qubit that provides information about errors occurring on ancilla qubits, which allows us to correct more errors in the subsequent QEC procedure.
We set weights for data errors and measurement errors based on conditional error probabilities conditioned on the measurement outcomes of local flag qubits.
Ref.~\cite{chamberland2020triangular} also uses flag information in the decoding graph, but there are a lot of differences from our method.
In Ref.~\cite{chamberland2020triangular}, flag edges are added and each weight is set by a conventional way, except for a renormalization for the weights of edges that are unlikely to flip, whereas we do not add additional nodes and edges.
We also propose a method to estimate the conditional error probabilities, which is efficient, accurate, and can be used even when the underlying noise is {\it a priori} unknown.
The conditional error probabilities are estimated by repeatedly executing a tailored quantum circuit offline prior to the decoding and obtaining information about errors from its measurement outcomes.
In addition, we perform deflagging procedure proposed in Refs.~\cite{PhysRevLett.128.110504,sundaresan2023demonstrating} combined with FWO to further improve the performance, which is a new application of deflagging.

In our syndrome measurement circuits using flag qubits, cat states are prepared first. The use of cat states reduces the circuit depth of syndrome circuits, and thus suppresses the occurrence of errors, as also mentioned in Ref.~\cite{graphmatching}.
At the same time, the number of data qubits interacting with an ancilla qubit decreases, thus the propagation of errors is suppressed.
As to the implementation on hardware, cat states reduce the connectivity requirement, which is especially essential in superconducting architectures.
As a decoder, we use the integer programming decoder, which is not efficient, throughout this paper.
This is because our focus is not on decoders themselves but on how their performance improves by the proposed method.
Although our method can be applied to other weight-based decoders, the simplicity of the formulation in the integer programming decoder makes the proposed method more straightforward to understand.
Our approach is versatile in that it can also be applied to other QECCs that have high-weight stabilizer generators, such as high-rate quantum low-density parity check (LDPC) codes \cite{breuckmann2021quantum}.

In numerical simulations, we assess logical error rates for memory errors and do not analyze time-like logical errors rates \cite{chamberland2022circuit}, which may be relevant for implementing non-Clifford gates.
We improve the threshold of the (4.8.8) color code under circuit-level noise from 0.14\% to around 0.27\%, which is calculated by using the integer programming decoder.
In addition, in the (6.6.6) color code, we achieved the circuit-level threshold of around 0.36\%, which is the same within statistical errors as the highest value of 0.37(1)\% \cite{PRXQuantum.2.020341}, among the previous studies that employ the same noise model.
Note that our threshold values are calculated by using code distances up to $d=9$.
Moreover, in both cases, an effective code distance is also improved compared to a conventional method that uses a single ancilla qubit for each stabilizer measurement and employs uniform weights, meaning that FWO helps correct large hook errors that arise from relatively few faults. Thereby, the achieved logical error rates at low physical error rates are almost one order of magnitude lower than those of the conventional method when comparing with the same code distance. We also verified that our method achieves lower logical error rates in most cases, even when comparing to the conventional method with a higher code distance, assuming a given number of available qubits where the conventional method with a higher code distance is feasible.

The rest of the paper is organized as follows.
In Sec.~\ref{preliminary}, we first introduce 2D color codes.
Then, we describe the decoding formulation and conventional weights of a decoder under circuit-level noise using the integer programming decoder.
We also explain conventional syndrome measurement circuits for color codes and the reasons why they lead to a low threshold.
In Sec.~\ref{proposed_method}, we describe the details of the procedure for the proposed FWO.
Additionally, we explain the deflagging procedure and how to estimate the conditional error probabilities required for setting the weights.
In Sec.~\ref{numerical simulation}, we show the numerical results for the logical error rates achieved by our method and discuss comparisons with existing methods.
Finally, a conclusion is presented in Sec.~\ref{conclusion}.
\section{Preliminaries}
\label{preliminary}
\subsection{Color codes}
\begin{figure}[t]
  \begin{center}
    \includegraphics[width=0.9\linewidth]{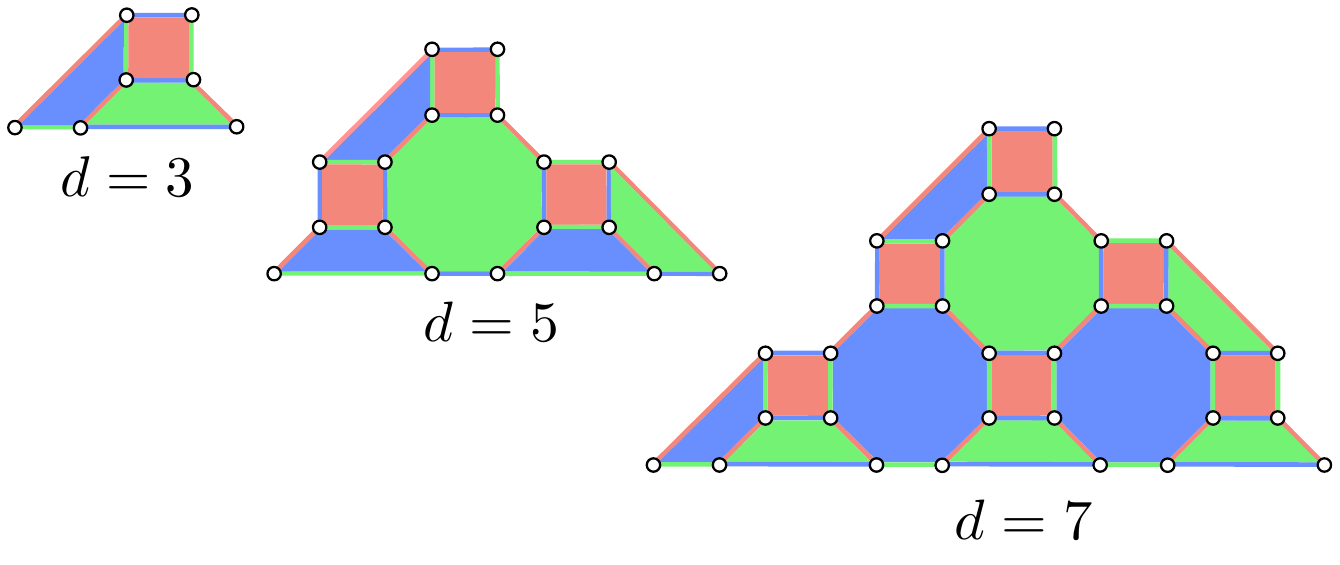}
  \end{center}
  \caption{The (4.8.8) color code for each code distance $d$. A white circle represents a data qubit.}
  \label{fig:square_octagonal_lattice}
\end{figure}

\begin{figure}[tb]
  \begin{center}
    \includegraphics[width=0.9\linewidth]{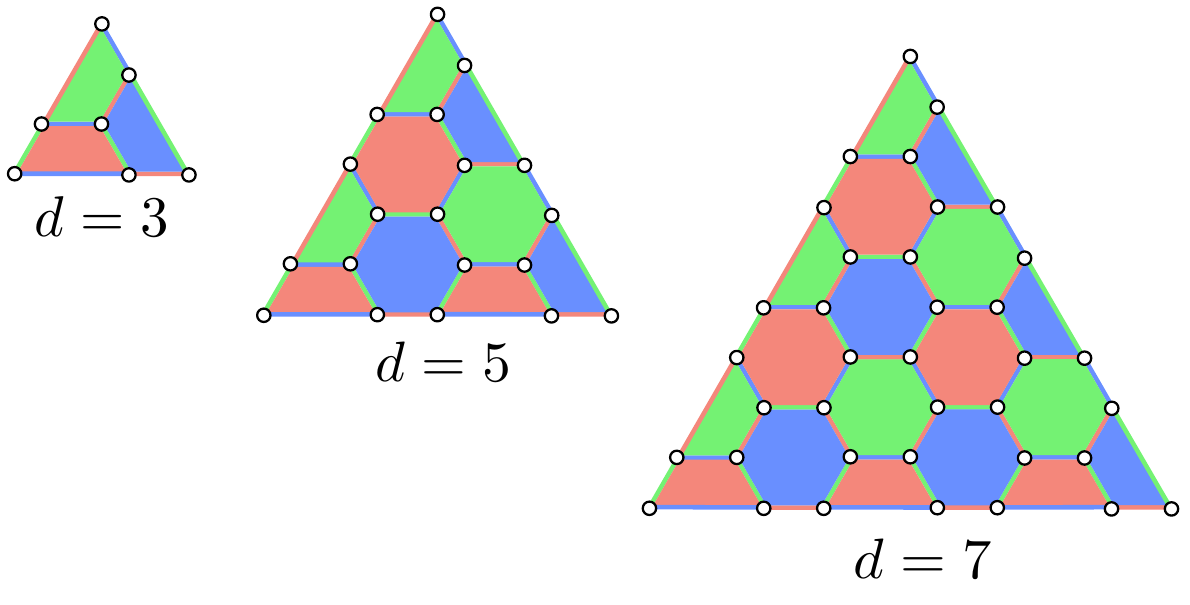}
  \end{center}
  \caption{The (6.6.6) color code for each code distance $d$.}
  \label{fig:hexagonal_lattice}
\end{figure}

Color codes are a type of topological code defined on a trivalent graph with three-colorable faces.
Each vertex $v$ of the graph corresponds to a qubit.
For each face $f$, $X$ and $Z$ stabilizer generators are defined as the tensor product of the Pauli $X$ and $Z$ operators acting on each qubit incident on the face, respectively:
\begin{equation}
  \label{colorxsta}
  G_{f}^{X}:=\prod_{v \in \delta f} X_v,
\end{equation}
\begin{equation}
  \label{colorzsta}
  G_{f}^{Z}:=\prod_{v \in \delta f} Z_v.
\end{equation}
Here, $\delta f$ is the set of vertices that touch $f$.
The code state is the simultaneous $+1$ eigenstate of all stabilizer operators.
There are three types of lattices that can be used to define 2D color codes: (4.8.8) lattice, (6.6.6) lattice, and (4.6.12) lattice \cite{landahl}.
Among these lattices, we focus on (4.8.8) lattice and (6.6.6) lattice in this work.
The (4.8.8) lattice is a semi-regular lattice where each vertex is incident to a square and two octagonal faces, and the (6.6.6) lattice is a regular lattice where each vertex is incident to three hexagonal faces.
The 2D color codes defined on the (4.8.8) lattice and the (6.6.6) lattice with boundaries, in which a single logical qubit is encoded, are shown in Figs.~\ref{fig:square_octagonal_lattice} and \ref{fig:hexagonal_lattice}, respectively.
In these 2D color codes, we can implement all logical Clifford gates transversally \cite{PhysRevA.91.032330}.
\subsection{Decoding under circuit-level noise using integer programming decoder}
\label{decoding_circuit_level}
Here, we explain the circuit-level noise model and how to decode under this noise model using the integer programming decoder.
The circuit-level noise model is a noise model that assumes every operation in a quantum circuit, including state preparations, gate operations, idling operations (i.e., when no operation is being performed), and measurements, can suffer from errors.
To define the circuit-level noise model, we first introduce a depolarizing channel.
The depolarizing channel is defined as
\begin{equation}
  \mathcal{E}_1(\rho_1) = (1 - p)\rho_1 + \frac{p}{3} \sum_{P \in \{X, Y, Z\}} P \rho_1 P,
\end{equation}
\begin{align}
  \mathcal{E}_2(\rho_2) = & (1 - p)\rho_2 + \frac{p}{15}                       \\ \nonumber
                          & \times \sum_{\substack{P_1, P_2 \in \{I, X, Y, Z\} \\ P_1 \otimes P_2 \neq I \otimes I}} (P_1 \otimes P_2) \rho_2 (P_1 \otimes P_2),
\end{align}
where $p$ is a physical error probability, $\rho_1$ is a density operator of a single qubit, and $\rho_2$ is a density operator of two qubits.
Then, the circuit-level noise model is defined as follows:
\begin{enumerate}[(i)]
  \item Each single-qubit gate (including identity gate) acts ideally, followed by the single-qubit depolarizing channel $\mathcal{E}_1$ with the probability $p$.
  \item Each two-qubit gate acts ideally, followed by the two-qubit depolarizing channel $\mathcal{E}_2$ with the probability $p$.
  \item Each state preparation fails with probability $p$, and an orthogonal state is prepared.
  \item Each measurement fails with probability $p$, and the measurement outcome is inverted.
\end{enumerate}

The task of a decoding is to estimate the most likely errors given a syndrome.
Under the circuit-level noise, a syndrome is not reliable due to the presence of measurement errors.
Thus, if a single syndrome measurement is performed and the decoding is carried out based on the syndrome, the decoding accuracy significantly gets worse.
In this situation, it is possible to suppress performance deterioration due to measurement errors by repeating the syndrome measurement multiple times (in this study, $d$ times).
Then, we decode $X$ and $Z$ errors separately based on the obtained syndrome in spacetime.
In the case of a noise model where probabilities of data errors and measurement errors occurring are assumed to be independently and identically distributed (i.i.d.), i.e., phenomenological noise model, we can estimate the most likely errors by minimizing the total number of errors given the syndrome.
However, under the circuit-level noise, the probabilities of cumulative data errors and measurement errors for each round, i.e., the probabilities that edges in the decoding graph are triggered, are not identical because the way errors occur and propagate differs for each qubit.
Also, they are not independent due to the correlated errors caused by entangling gates.
Therefore, to identify the most likely errors, it is necessary to estimate the errors based on an actual error probability distribution.

In the following, we explain how to decode under the circuit-level noise using the integer programming decoder, where the decoding problem is formulated as an integer programming problem.
The decoding formulation is an extension of that in the phenomenological noise model provided in Ref.~\cite{PhysRevResearch.6.013092}.
Here, we explain only the decoding of $X$ errors, but the decoding of $Z$ errors is also performed using the same algorithm.
In this decoder, errors and syndrome values are represented as binary variables.
Here, we define the binary variable $u^{(t)}_v \in \{0,1\}$ as the cumulative data error that has occurred on the qubit at vertex $v$ until time $t$, and the binary variable $r^{(t)}_f \in \{0,1 \}$ as the measurement error occurred on the face $f$ at time $t$.
Then, we can express the measured syndrome value on the face $f$ at time $t$ as the binary value $s^{(t)}_f \in \{0,1\}$:
\begin{equation}
  \label{sft}
  s^{(t)}_f=\bigoplus_{v \in \delta f} u^{(t)}_v \oplus r^{(t)}_f.
\end{equation}
The syndrome difference between time $t$ and time $t-1$, i.e., $s^{(t)}_f \oplus s^{(t-1)}_f$ detects the data errors occurred at time $t$.
Thus, the equation that the syndrome values should satisfy is given as follows:
\begin{equation}
  \label{ip_const}
  \bigoplus_{v \in \delta f} x^{(t)}_v \oplus r^{(t)}_f \oplus  r^{(t-1)}_f= s^{(t)}_f \oplus s^{(t-1)}_f \quad \forall f,
\end{equation}
where the binary variable $x^{(t)}_v \in \{0,1\}$ denotes newly occuring data error at time $t$, which corresponds to the XOR of $u^{(t)}_v$ and $u^{(t-1)}_v$.
If $x^{(t)}_v = 1$, it indicates that a data error has newly occurred on the corresponding data qubit at time $t$. Conversely, if $x^{(t)}_v = 0$, it indicates that a data error has not newly occurred on the corresponding data qubit at time $t$.
By setting weights for each error and minimizing the weighted number of errors satisfying Eq.~\eqref{ip_const}, it is possible to decode based on an error probability distribution.
Thus, the decoding problem under the circuit-level noise is formulated as an integer programming problem as follows:
\begin{equation}
  \mathrm{min} \quad \sum _{v,t}w_v^{(t)}x_v^{(t)} +\sum _{f,t}w_f^{(t)}r_f^{(t)},
  \label{ip_circuit_level_cost}
\end{equation}
\begin{equation}
  \label{ip_circuit_level_sto}
  \mathrm{s.t.} \quad   \bigoplus_{v \in \delta f} x^{(t)}_v \oplus r^{(t)}_f \oplus  r^{(t-1)}_f= s^{(t)}_f \oplus s^{(t-1)}_f \quad \forall f.
\end{equation}
Here, $w_v^{(t)}$ and $w_f^{(t)}$ are weights for data errors and measurement errors, respectively. Conventionally, they are defined as follows \cite{planar}:
\begin{equation}
  w^{(t)}_v= - \log \frac{p\left(x^{(t)}_v=1\right)}{1-p\left(x^{(t)}_v=1\right)},
  \label{weight_data_previous}
\end{equation}
\begin{equation}
  w^{(t)}_f= - \log \frac{p\left(r^{(t)}_f=1 \right)}{1-p\left(r^{(t)}_f=1 \right)}.
  \label{weight_meas_previous}
\end{equation}
For the details on the definition of the weights, see Appendix \ref{appendix:a}.

Note that an error probability distribution used in Eqs.~\eqref{weight_data_previous} and \eqref{weight_meas_previous} needs to be estimated.
So far, several methods have been proposed to estimate the error probability distribution in this context.
Refs.~ \cite{surface_circuit_threshold5,sundaresan2023demonstrating,PhysRevX.10.011022,zhang2023facilitating} have proposed methods to estimate the error probability distribution by analytically counting possible error events.
In Refs.~\cite{spitz2018adaptive,google2021exponential}, the error probability distribution is estimated by calculating the expected values of the syndrome from appropriate syndrome data.
In Ref.~\cite{wang2023dgr}, the error probability distribution is estimated by repeatedly decoding the obtained syndrome.
\subsection{Conventional syndrome measurement circuit}
\begin{figure}[b]
  \centering
  \includegraphics[width=\linewidth]{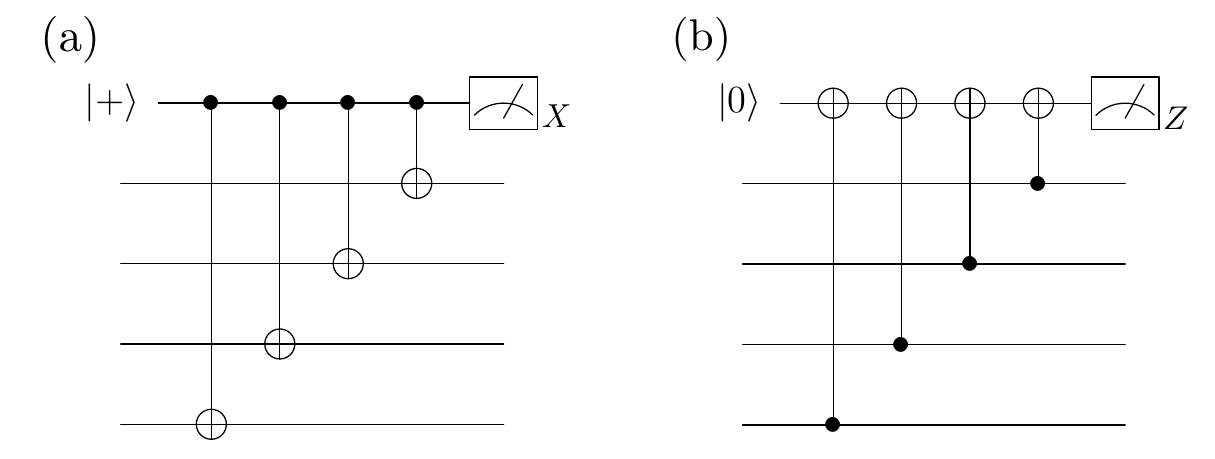}
  \caption{Conventional syndrome measurement circuits. (a) Circuit for measuring $X^{\otimes 4}$. (b) Circuit for measuring $Z^{\otimes 4}$.}
  \label{fig:syndrome_measurement_circuit}
\end{figure}
\begin{figure}[t]
  \centering
  \includegraphics[width=0.52\linewidth]{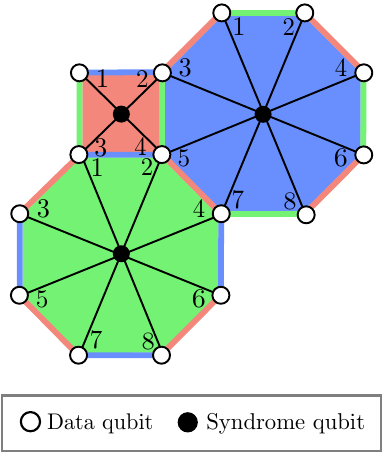}
  \caption{CNOT order for the (4.8.8) color code with a single ancilla qubit for each face. The black numbers indicate the time steps in which the CNOT gates are applied. The black lines represent interactions between qubits. The total number of qubits required for this method is $n_{4.8.8}^{\text{single}}(d)=(3d^2+6d-5)/4$.}
  \label{fig:square_octagonal_order_oneancilla}
\end{figure}
\begin{figure}[h]
  \centering
  \includegraphics[width=0.52\linewidth]{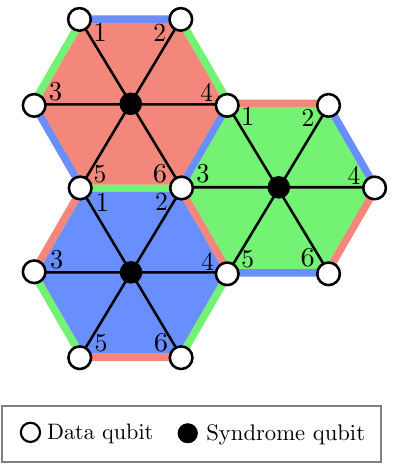}
  \caption{CNOT order for the (6.6.6) color code with a single ancilla qubit for each face. The total number of qubits required for this method is $n_{6.6.6}^{\text{single}}(d)=(9d^2-1)/8$.}
  \label{fig:hexagonal_order_oneancilla}
\end{figure}
\begin{figure}[tb]
  \centering
  \includegraphics[width=0.95\linewidth]{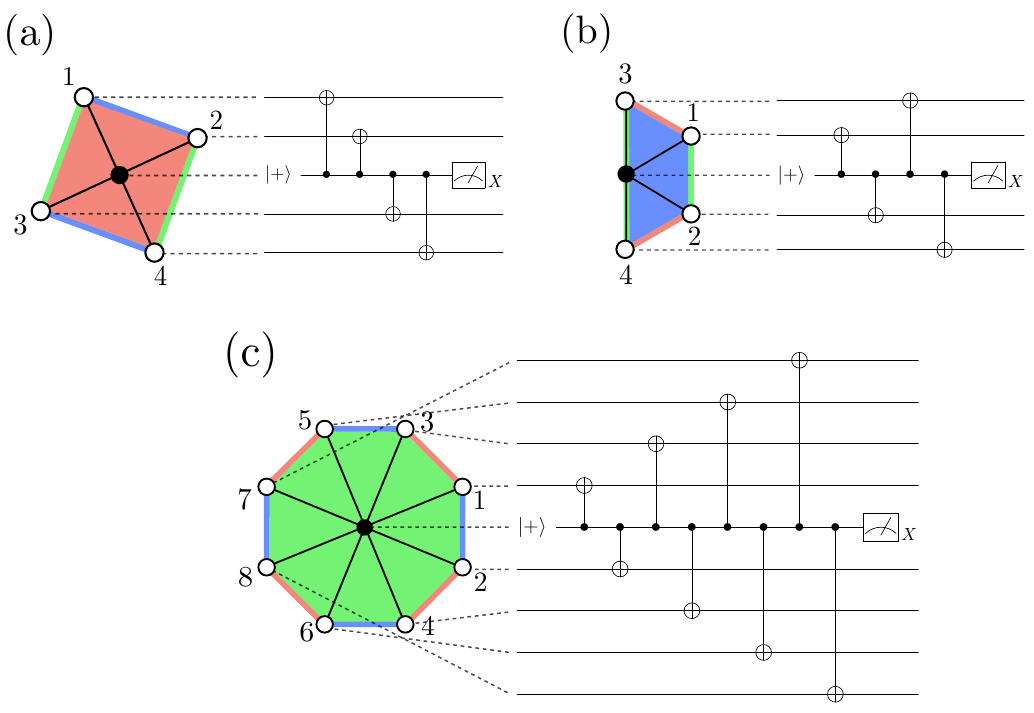}
  \caption{$X$ stabilizer measurement circuits of the (4.8.8) color code with a single ancilla qubit for each face. (a) Square face. (b) Trapezoidal face. (c) Octagonal face.}
  \label{fig:square_octagonal_circuit_oneancilla}
\end{figure}
\begin{figure}[tb]
  \centering
  \includegraphics[width=0.95\linewidth]{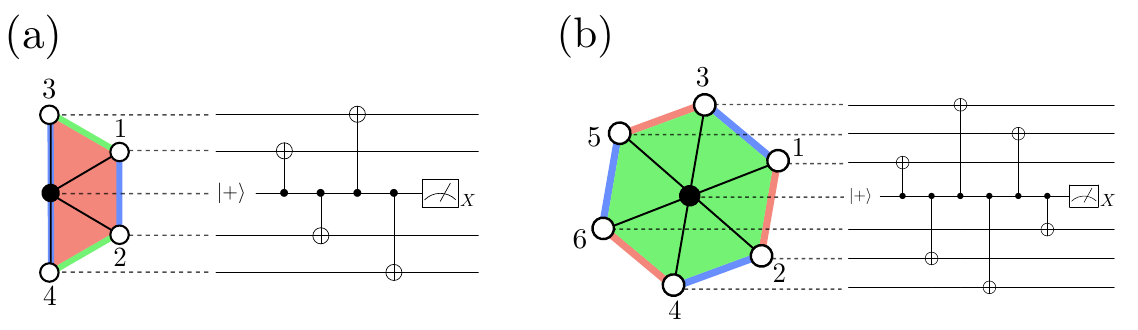}
  \caption{$X$ stabilizer measurement circuits of the (6.6.6) color code with a single ancilla qubit for each face. (a) Trapezoidal face. (b) Hexagonal face.}
  \label{fig:hexagonal_circuit_oneancilla}
\end{figure}

Syndrome measurements need to be performed without destroying the encoded state.
If data qubits are directly measured, the superposition state will be destroyed, so indirect measurements are used for syndrome measurements.
In the conventional method \cite{surface_circuit_threshold2,landahl}, measurements of $X$ and $Z$ stabilizer generators are performed using a single ancilla qubit, as shown in Figs.~\ref{fig:syndrome_measurement_circuit}(\hyperref[fig:syndrome_measurement_circuit]{a}) and \ref{fig:syndrome_measurement_circuit}(\hyperref[fig:syndrome_measurement_circuit]{b}), respectively.
In this method, the more data qubits the stabilizer generators act on, the deeper the circuit depth of the syndrome measurement circuit, resulting in more locations where errors may occur.
Also, the more data qubits interact with an ancilla qubit, the more error propagation paths there are, so that an error can propagate widely.
Here, an error propagation path is a set of locations in a circuit where an occurring error propagates.
Thus, color codes with high-weight stabilizer generators are more susceptible to errors compared to surface codes, which have at most weight-four stabilizer generators.
This is the reason why the thresholds of color codes are low under the circuit-level noise.
Here, we consider a CNOT schedule where the $X$ stabilizer measurement is performed first, followed by the $Z$ stabilizer measurement.
In Ref.~\cite{landahl}, it has been shown that the threshold of the (4.8.8) color code using this syndrome measurement circuit is around 0.08\%.
Examples of a possible CNOT order are shown for the (4.8.8) color code in Fig.~\ref{fig:square_octagonal_order_oneancilla} and for the (6.6.6) color code in Fig.~\ref{fig:hexagonal_order_oneancilla}.
The total number of required qubits, including both data and ancilla qubits, is $n_{4.8.8}^{\text{single}}(d)=(3d^2+6d-5)/4$ for the (4.8.8) color code and $n_{6.6.6}^{\text{single}}(d)=(9d^2-1)/8$ for the (6.6.6) color code.
We also show the $X$ stabilizer measurement circuits for each face of the (4.8.8) color code in Fig.~\ref{fig:square_octagonal_circuit_oneancilla} and of the (6.6.6) color code in Fig.~\ref{fig:hexagonal_circuit_oneancilla}.
$Z$ stabilizer measurement circuits are similar, except that the basis changes to the $Z$-basis, and the direction of the CNOT gates is reversed.
We need to implement CNOT gates with a depth of 8 for the (4.8.8) color code and a depth of 6 for the (6.6.6) color code.
\section{Improving threshold for fault-tolerant color code quantum computing}
\label{proposed_method}
Using conventional syndrome measurement circuits and weights results in a lower threshold in color codes due to their high-weight stabilizer generators.
Here, we propose a method to improve the thresholds of color codes under the circuit-level noise.
In the following, we first describe the syndrome measurement circuit employed here.
Subsequently, we explain how to set the weights of decoders.
The deflagging procedure, which is carried out to further improve the performance, is also introduced.
We also propose a method for estimating conditional error probabilities, which is also one of our main contributions.
\subsection{Syndrome measurement gadget}
\begin{figure}[b]
  \centering
  \includegraphics[width=0.9\linewidth]{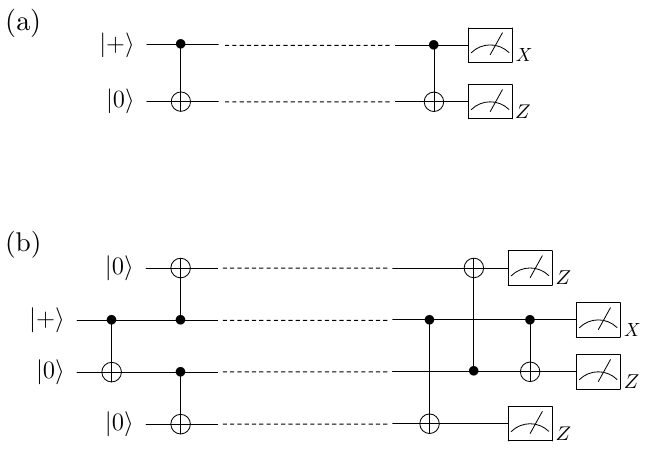}
  \caption{Flag gadgets for $X$ stabilizer measurements. (a) two-qubit flag gadget. (b) four-qubit flag gadget. The dotted lines in the circuits represent the interaction with the data qubits. The $X$-basis measurement provides the syndrome value, and the $Z$-basis measurements provide the flag values.}
  \label{fig:flag_gadget}
\end{figure}

\begin{figure}[tb]
  \centering
  \includegraphics[width=0.75\linewidth]{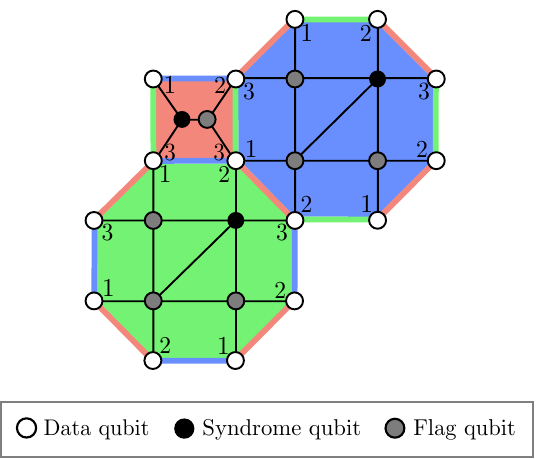}
  \caption{CNOT order for the (4.8.8) color code with the flag gadgets. The total number of qubits required for this method is $n_{4.8.8}^{\text{FWO}}(d)=(5d^2+4d-5)/4$.}
  \label{fig:square_octagonal_order}
\end{figure}

\begin{figure}[t]
  \centering
  \includegraphics[width=0.75\linewidth]{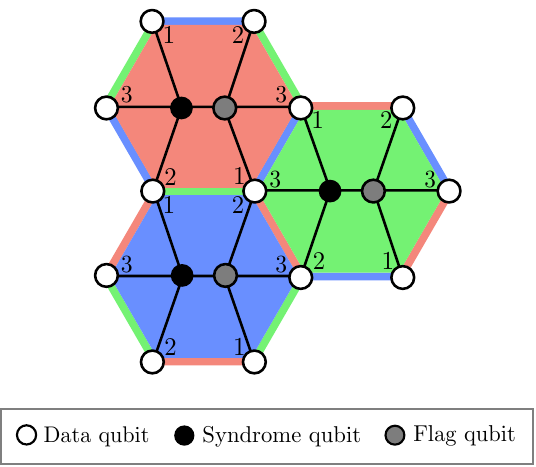}
  \caption{CNOT order for the (6.6.6) color code with the flag gadgets. The total number of qubits required for this method is $n_{6.6.6}^{\text{FWO}}(d)=(3d^2-1)/2$.}
  \label{fig:hexagonal_order}
\end{figure}

\begin{figure}[ht]
  \centering
  \includegraphics[width=0.95\linewidth]{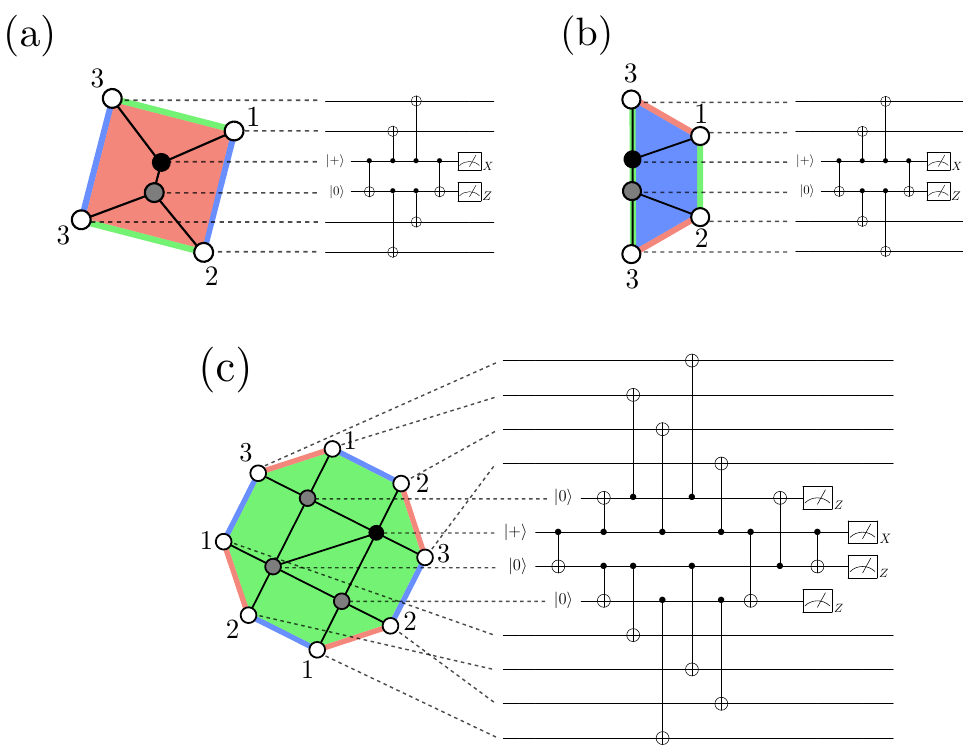}
  \caption{$X$ stabilizer measurement circuits of the (4.8.8) color code with the flag gadgets. (a) Square face. (b) Trapezoidal face. (c) Octagonal face.}
  \label{fig:square_octagonal_circuit}
\end{figure}

\begin{figure}[t]
  \centering
  \includegraphics[width=0.95\linewidth]{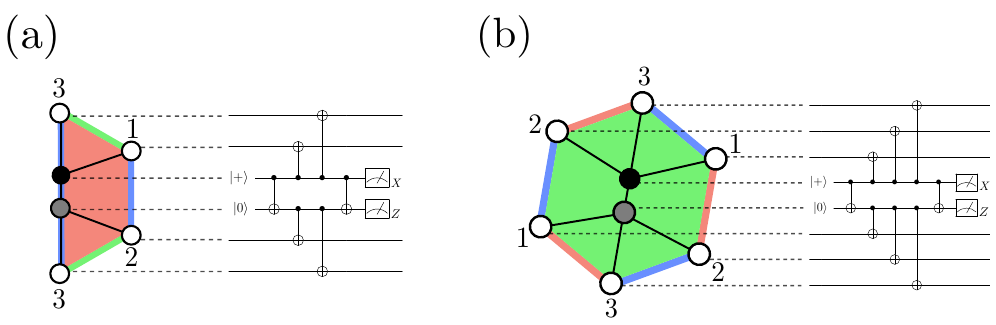}
  \caption{$X$ stabilizer measurement circuits of the (6.6.6) color code with the flag gadgets. (a) Trapezoidal face. (b) Hexagonal face.}
  \label{fig:hexagonal_circuit}
\end{figure}

In a syndrome measurement, we use a {\it flag gadget} for each face to extract a syndrome instead of using a single ancilla qubit for each face.
Fig.~\ref{fig:flag_gadget} shows two types of flag gadgets used for $X$ stabilizer measurements: a {\it two-qubit flag gadget} \cite{baireuther2019neural} and a {\it four-qubit flag gadget} \cite{PhysRevLett.98.020501}.
In these gadgets, the qubits prepared in $\ket{+}$ act as syndrome qubits, while those prepared in $\ket{0}$ act as flag qubits.
Flag gadgets for $Z$ stabilizer measurements are obtained by reversing the basis and the direction of the CNOT gates.
These gadgets are circuits that can flag certain types of errors from the measurement outcomes and, at the same time, provide the syndrome.
They first prepare a cat state, which is defined as follows:
\begin{equation}
  \ket{\text{cat}}=\frac{\ket{0} ^{\otimes n} + \ket{1} ^{\otimes n}}{\sqrt{2}}.
\end{equation}
After preparing the cat state, they interact with data qubits and finally perform post-processing and measurements.
The cat state allows us to parallelize CNOT gates and reduce the depth for the syndrome measurement, resulting in less idling noise.
Additionally, a decrease in the error propagation paths leads to a reduction in the error propagation.
For the (4.8.8) color code, the four-qubit flag gadget is used for the weight-8 stabilizer measurements, and the two-qubit flag gadget is used for the weight-four stabilizer measurements, the same as in Ref.~\cite{graphmatching}.
We show the CNOT order for the (4.8.8) color code when using the flag gadgets in Fig.~\ref{fig:square_octagonal_order}.
The CNOT gates are applied to the two neighboring data qubits from each qubit in the gadget.
For the (6.6.6) color code, the two-qubit flag gadget is used for each stabilizer measurement.
The CNOT order for the (6.6.6) color code is shown in Fig.~\ref{fig:hexagonal_order}.
The CNOT gates are applied to the three neighboring data qubits from each qubit in the gadget for the weight-6 stabilizer measurements and to the two neighboring data qubits for the weight-four stabilizer measurements.
Since an ancilla qubit interacts with only two or three data qubits, the propagation of errors from the ancilla qubits to the data qubits is suppressed compared to the conventional syndrome measurement circuit.
Furthermore, the depth of the CNOT gates acting on the data qubits has been reduced to three steps in both lattices, which is the minimum in the case of color codes.
This is because in color codes, up to three stabilizer generators are involved for a data qubit, and at each time, at most one operation can be applied to any qubit.
The total number of data and ancilla qubits required for this way of measuring syndrome is $n_{4.8.8}^{\text{FWO}}(d)=(5d^2+4d-5)/4$ for the (4.8.8) color code and $n_{6.6.6}^{\text{FWO}}(d)=(3d^2-1)/2$ for the (6.6.6) color code.
The $X$ stabilizer measurement circuits for each face of the (4.8.8) color code and the (6.6.6) color code are shown in Figs.~\ref{fig:square_octagonal_circuit} and \ref{fig:hexagonal_circuit}, respectively.
\subsection{Flagged weight optimization}
\begin{figure}[tb]
  \centering
  \includegraphics[width=0.9\linewidth]{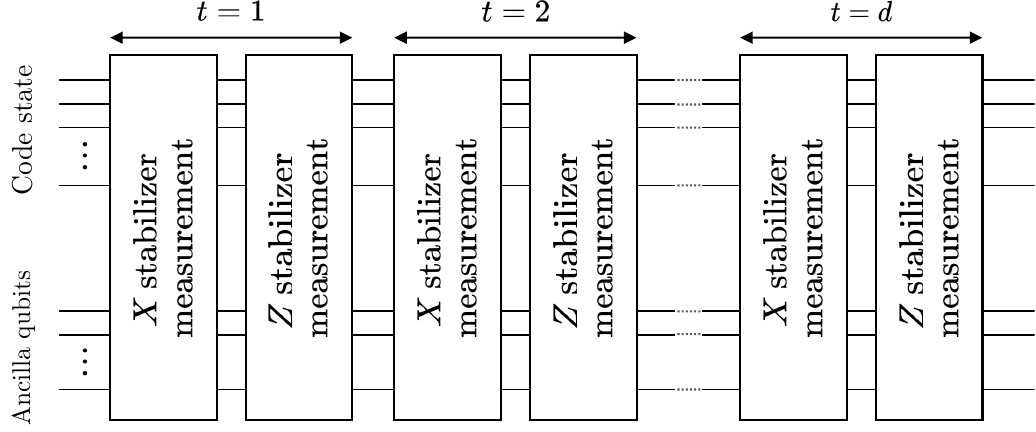}
  \caption{Syndrome measurement procedure.}
  \label{fig:syndrome_meas_procedure}
\end{figure}

We explain the setting of the decoder weights.
Here, each cycle of a syndrome measurement consists of performing an $X$ stabilizer measurement followed by a $Z$ stabilizer measurement, as shown in Fig.~\ref{fig:syndrome_meas_procedure}.
Let us consider the case of performing $X$ error correction.
If an $X$ error occurring on an ancilla qubit for an $X$ stabilizer measurement propagates to data qubits as a hook error, it is detected in the subsequent $Z$ stabilizer measurement.
Thus, the flag values of the $X$ stabilizer measurement at time $t$ provide information about the $X$ data errors at time $t$.
Note that flag values are not perfect, as errors can also occur on the flag qubits.
Here, in order to incorporate the flag information into the decoder weights defined by Eqs.~\eqref{weight_data_previous} and \eqref{weight_meas_previous}, we set the weights using conditional error probabilities conditioned on the flag values.
For weights of edges corresponding to $X$ data errors, we use the conditional error probabilities conditioned on all flag values in the measurements of the $X$ stabilizers acting on the corresponding data qubits at the same time step:
\begin{equation}
  w^{(t)}_v= - \log \frac{p\left(x^{(t)}_v=1 |\bigcup_{f:v \in \delta f} \mathcal{F}^{(t)}_{f,X} \right)}{1-p\left(x^{(t)}_v=1 |\bigcup_{f:v \in \delta f} \mathcal{F}^{(t)}_{f,X}\right)}.
  \label{weight_data_x_error_correction}
\end{equation}
Here, $\mathcal{F}^{(t)}_{f,\sigma}$ is a set of flag values in the flag gadget used for measuring the $\sigma \in \{X,Z\}$ stabilizers defined on the face $f$ at time $t$.
Flag values of a $Z$ stabilizer measurement explicitly provide information only about $Z$ errors occurred in the ancilla qubits, but they also implicitly provide information about $X$ errors occurred in the ancilla qubits.
To elaborate, a trivial (i.e., unflipped) flag value implies that either an $X$ error that is not correlated with a $Z$ error occurred or no error occurred at a certain location.
Therefore, for weights of edges corresponding to measurement errors, we use the conditional error probabilities conditioned on the flag values in the measurements of the $Z$ stabilizers that measure the corresponding syndrome:
\begin{equation}
  w^{(t)}_f= - \log \frac{p\left(r^{(t)}_f=1 |\mathcal{F}^{(t)}_{f,Z} \right)}{1-p\left(r^{(t)}_f=1 |\mathcal{F}^{(t)}_{f,Z} \right)}.
  \label{weight_meas_x_error_correction}
\end{equation}
On the other hand, in the case of performing $Z$ error correction, the weights of edges corresponding to $Z$ data errors and measurement errors are set as follows:
\begin{equation}
  w^{(t)}_v=
  \begin{cases}
    - \log \frac{p\left(z^{(1)}_v=1 \right)}{1-p\left(z^{(1)}_v=1 \right)}                                                                                             & \text{if } t  = 1, \\
    - \log \frac{p\left(z^{(t)}_v=1 |\bigcup_{f:v \in \delta f} \mathcal{F}^{(t-1)}_{f,Z}\right)}{1-p\left(z^{(t)}_v=1 |\bigcup_{f:v \in \delta f} \mathcal{F}^{(t-1)}_{f,Z}\right)} & \text{otherwise.},
  \end{cases}
  \label{weight_data_z_error_correction}
\end{equation}
\begin{equation}
  w^{(t)}_f= - \log \frac{p\left(r^{(t)}_f=1 |\mathcal{F}^{(t)}_{f,X} \right)}{1-p\left(r^{(t)}_f=1 |\mathcal{F}^{(t)}_{f,X} \right)}.
  \label{weight_meas_z_error_correction}
\end{equation}
In Eq.~\eqref{weight_data_z_error_correction}, we do not use flag information for the weights of the edges corresponding to $Z$ data errors at the initial time step.
This is because the procedure for a syndrome measurement consists of a cycle of first performing an $X$ stabilizer measurement, followed by a $Z$ stabilizer measurement, so there is no flag information for the $Z$ data errors at the initial time step.

The memory requirement to store the weights scales linearly with respect to the number of data qubits because each probability of data error or measurement error occurring is conditioned only on measurement outcomes of locally located flag qubits, the number of which is constant.
When applying this method to a decoder, the increase in computational overhead for the decoding is solely due to the computational cost of retrieving weight values from stored information based on the flag values.
This overhead is similar to that of a common weight-setting procedure.
\subsection{Deflagging}
In addition to FWO, we perform a deflagging procedure proposed in Refs.~\cite{PhysRevLett.128.110504,sundaresan2023demonstrating}, i.e., immediately apply Pauli operators corresponding to the errors that are implied by the flag values to the data qubits.
Without directly applying the Pauli operators to the data qubits, we can perform the same operation by just recording the Pauli operators as classical information and updating the Pauli frame \cite{knill2005quantum, Aliferis2005QuantumAT}.
Deflagging is also performed when estimating the conditional error probabilities.
This deflagging procedure enables us to correct some errors that cannot be corrected with only FWO.
For the details of the deflagging procedure and how it improves logical error rates, see Appendix \ref{appendix:b}.
\subsection{Estimating conditional error probabilities}
\begin{figure*}[t]
  \centering
  \includegraphics[width=\linewidth]{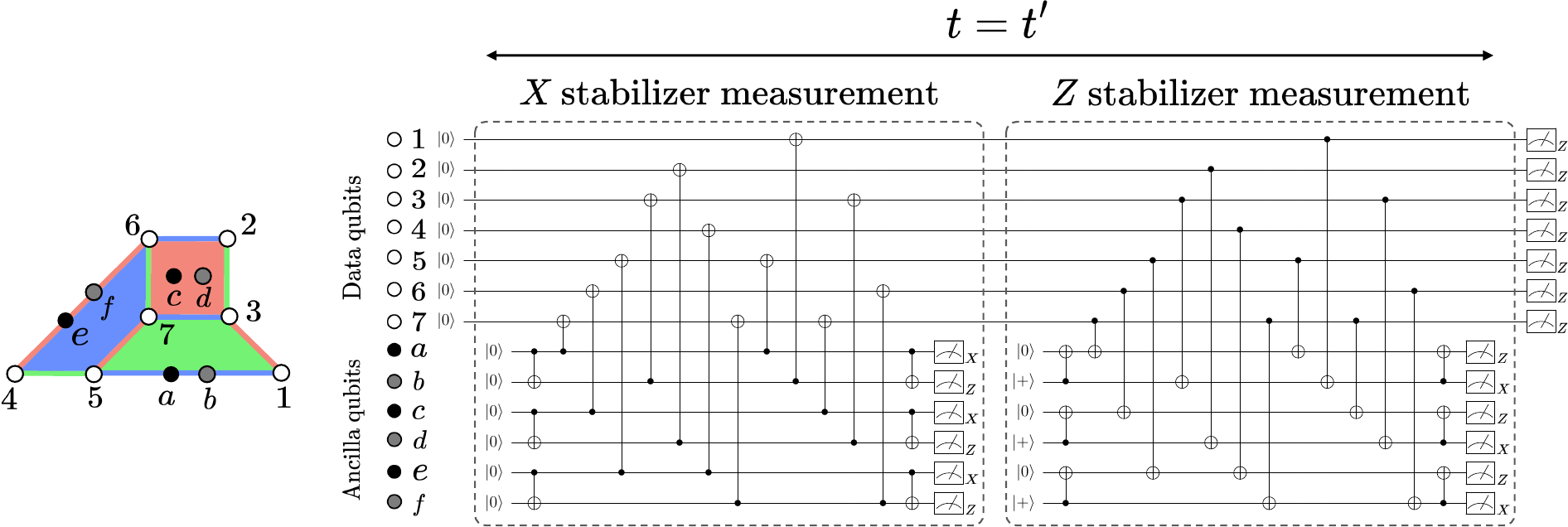}
  \caption{The quantum circuit $C_X$ for the distance-three (4.8.8) color code. This circuit is tailored for estimating the conditional error probabilities used in the weights in the $X$ error correction decoder.}
  \label{fig:conditional_prob_est_circuit}
\end{figure*}

The conditional error probabilities in Eqs.~\eqref{weight_data_x_error_correction}-\eqref{weight_meas_z_error_correction} need to be estimated before decoding.
Here, we propose a method to estimate the conditional error probabilities that is accurate, efficient, and can be used even when the underlying noise is {\it a priori} unknown.
We estimate the conditional error probabilities by running a tailored quantum circuit multiple times.
We employ quantum circuits $C_X$ and $C_Z$, each tailored for estimating the conditional error probabilities used in the edge weights in the $X$ and $Z$ error correction decoders, respectively.
These quantum circuits are modified versions of a one-cycle syndrome measurement circuit, where the initial states of certain qubits are altered and the data qubits are measured transversally at the end.
In $C_X$, an $X$ stabilizer measurement circuit is executed with all data qubits and ancilla qubits initialized to $\ket{0}$, followed by a $Z$ stabilizer measurement.
Then, the data qubits are measured transversally in the $Z$-basis.
In the case of $C_Z$, we use a one-cycle syndrome measurement circuit, where a $Z$ stabilizer measurement is performed, followed by an $X$ stabilizer measurement.
The initial states of all data qubits and ancilla qubits used for the $Z$ stabilizer measurement are prepared as $\ket{+}$.
At the end, the data qubits are measured in the $X$-basis in a transversal manner.
These circuits enable a {\it direct} detection of data errors and also allow us to estimate measurement errors that can occur in the circuits.
Here, a direct detection means detecting errors occurring in a qubit solely based on the outcome of single-qubit measurement for that qubit.
We show $C_X$ for the case of the distance-three (4.8.8) color code in Fig.~\ref{fig:conditional_prob_est_circuit}.
The following describes the procedure for estimating the conditional error probabilities using $C_X$.
\begin{enumerate}[i.]
  \item Execute $C_X$ with deflagging.
  \item Record the data errors and the flag values from the measurement outcomes of the data qubits and the flag qubits, respectively.
  \item By taking XOR of the proper combination of the obtained data errors, calculate the ideal syndrome that the data errors should give. Compare this ideal syndrome with the syndrome obtained from the measurement outcomes of the syndrome qubits. If there are syndrome values that are inconsistent between the two, record that measurement errors have occurred in the corresponding syndrome values.
  \item Repeat the steps up to this point a sufficiently large number of times. Then, to estimate the probability of a data or measurement error given an observed set of flag outcomes, divide the total number of times each error was observed with the set of flag outcomes, by the total number of times the set of flag outcomes was observed.
\end{enumerate}
Since the syndrome measurement is repetitive, the conditional error probabilities estimated by this procedure is used at all time steps.
Note that in $C_X$ and $C_Z$, preparing the initial state in this way is necessary for the direct detection of data errors.
This is because without such a modification of the initial states, each measurement outcome of the transversal measurement for data qubits at the end of the syndrome measurement circuit cannot be used for the direct detection of data errors.
The reason is that, when detecting $X$ ($Z$) errors, the data qubits and the ancilla qubits used in the $X$ ($Z$) stabilizer measurement circuits are in a superposition state.
Once the conditional error probabilities are estimated offline, those values can be used in all subsequent decoding.
\section{Numerical simulation}
\label{numerical simulation}
\subsection{Settings}
We perform Monte Carlo simulations to compute the logical error rates when the proposed method is applied to the integer programming decoder under the circuit-level noise.
The number of samples in the Monte Carlo simulations is $10^6$.
In order to evaluate logical error rates, we need to compare the input logical state with the output logical state. Since the quantum state after the decoding and recovery operation in space-time may not necessarily be in the code space, we perform an ideal error correction at the final time-slice \cite{Aliferis2005QuantumAT,landahl} to project the state onto the code space.
For comparison, we also compute the logical error rates of the existing method that uses a single ancilla qubit for each syndrome measurement and employs uniform decoder weights.
We call this existing method the {\it single ancilla method} in the following.
We use Stim \cite{gidney2021stim} to implement the quantum circuits and CPLEX \cite{cplex} for the integer programming solver.
The source code we used is available
on GitHub \cite{github_code}.

FTQC is thought to be performed at a physical error rate far below the threshold, so the behavior of the logical error rates in the low physical error rate region is essential.
In the region where the physical error rate is sufficiently low, the logical error rate $p_{\mathrm{L}}$ scales as follows \cite{surface_circuit_threshold2}:
\begin{equation}
  p_{\mathrm{L}}=c\left( \frac{p}{p_\mathrm{th}^*} \right)^{\alpha\left(\frac{d+1}{2} \right)}.
  \label{fitting_low_error_rate}
\end{equation}
Here, $c$ is a constant, $p$ is a physical error rate, $p_\mathrm{th}^*$ is a threshold, $d$ is a code distance, and $\alpha$ is a constant indicating the effective code distance.
We refer to the threshold $p_\mathrm{th}^*$ obtained from Eq.~\eqref{fitting_low_error_rate} as the {\it scaling threshold}.
We fit the logical error rates using Eq.~\eqref{fitting_low_error_rate} with $c$, $p_\mathrm{th}^*$, and $\alpha$ as fitting parameters.

We also calculate the threshold obtained from the crossing point of data for different code distances. 
For a sufficiently large code distance $d$, the logical error rates $p_L$ around the threshold under circuit-level noise are expected to behave as follows \cite{landahl,scaling2}:
\begin{equation}
  p_{\mathrm{L}}=A+B(p-p_\mathrm{th}^\times)d^{1/\nu_0}+C(p-p_\mathrm{th}^\times)^2d^{2/\nu_0}.
  \label{fitting_near_threshold}
\end{equation}
Here, $A$, $B$, and $C$ are constants, $p$ is a physical error rate, $p_\mathrm{th}^\times$ is a threshold, $d$ is a code distance, and $\nu_0$ is a critical exponent.
We refer to the threshold $p_\mathrm{th}^\times$ obtained from Eq.~\eqref{fitting_near_threshold} as the {\it cross threshold}.
We fit the logical error rates using Eq.~\eqref{fitting_near_threshold} with $A$, $B$, $C$, $p_\mathrm{th}^\times$, and $\nu_0$ as fitting parameters.
\subsection{Results}
\begin{figure*}[t]
  \centering
  \includegraphics[width=0.95\linewidth]{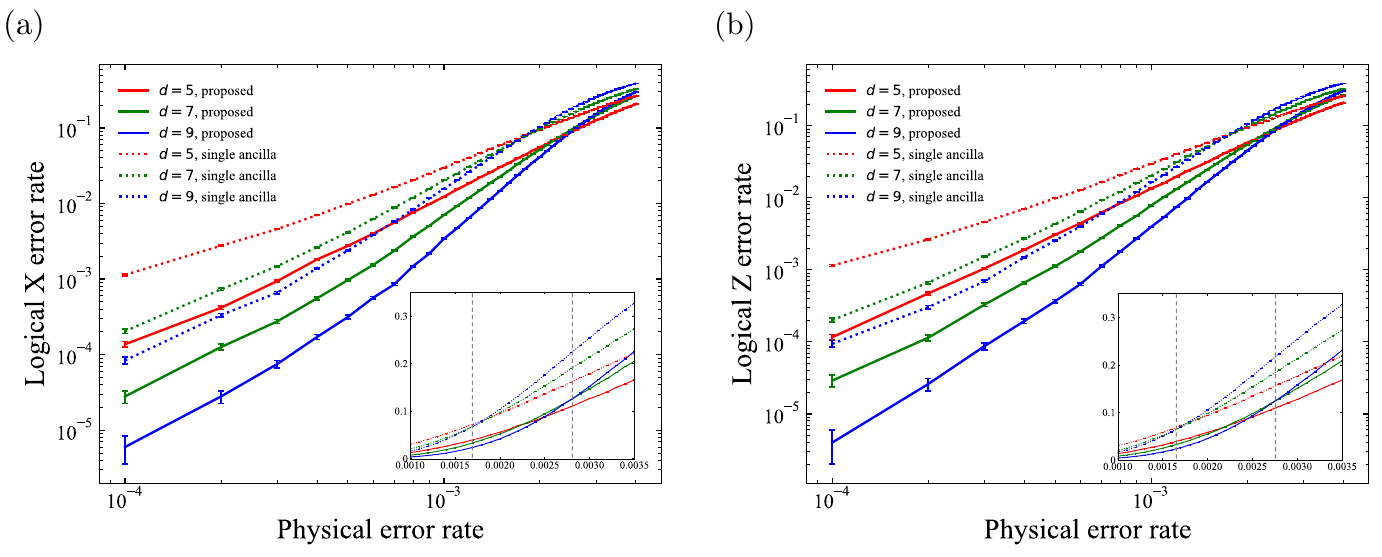}
  \caption{Logical error rates of the (4.8.8) color code for various code distances. The syndrome measurements are performed for $d$ rounds. The cross thresholds indicated by the gray vertical dashed lines in the figures are calculated by using the data of $d=7$ and $d=9$. (a) Logical $X$ error rates. The scaling threshold and cross threshold for the proposed method (solid lines) are $p_\mathrm{th}^*=0.29(1)\%$ and $p_\mathrm{th}^\times=0.279(1)\%$, respectively. The thresholds for the single ancilla method (dotted lines) are $p_\mathrm{th}^*=0.16(1)\%$ and $p_\mathrm{th}^\times=0.1694(5)\%$. (b) Logical $Z$ error rates. The thresholds for the proposed method (solid lines) are $p_\mathrm{th}^*=0.268(9)\%$ and $p_\mathrm{th}^\times=0.276(1)\%$. The thresholds for the single ancilla method (dotted lines) are $p_\mathrm{th}^*=0.14(1)\%$ and $p_\mathrm{th}^\times=0.164(2)\%$.}
  \label{fig:square_octagonal_error_rate}
\end{figure*}
\begin{figure*}[t]
  \centering
  \includegraphics[width=0.95\linewidth]{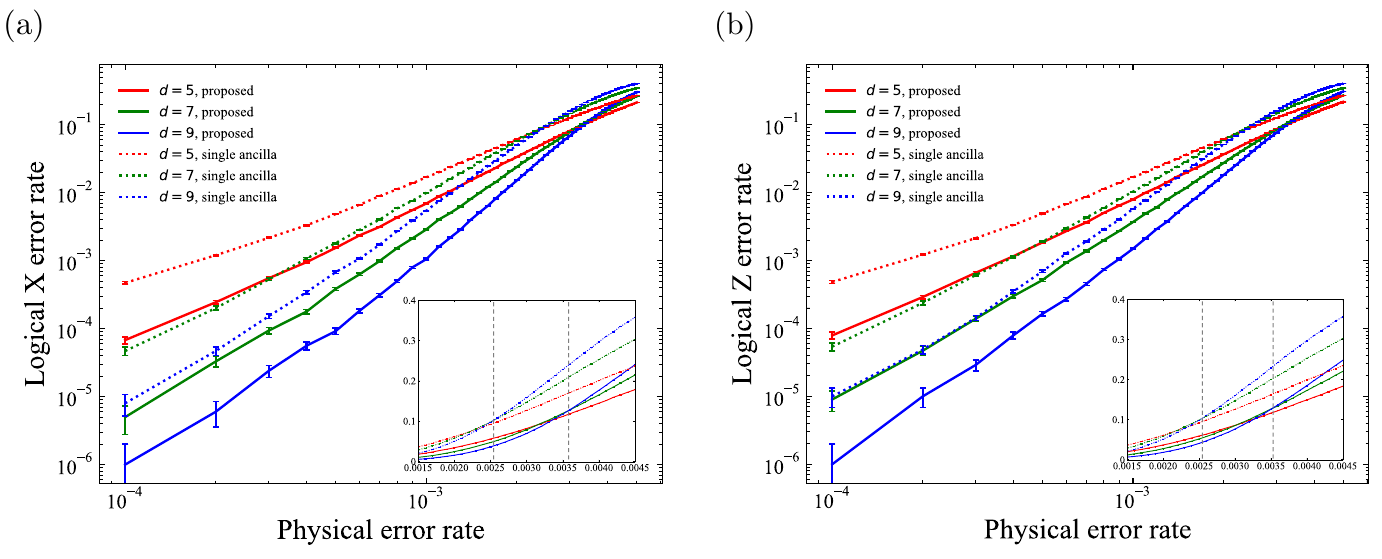}
  \caption{Logical error rates of the (6.6.6) color code for various code distances. The syndrome measurements are repeated $d$ times. The gray vertical dashed lines indicate the cross thresholds, which are calculated by using the data of $d=7$ and $d=9$. (a) Logical $X$ error rates. The scaling threshold and cross threshold for the proposed method (solid lines) are $p_\mathrm{th}^*=0.37(1)\%$ and $p_\mathrm{th}^\times=0.3575(6)\%$, respectively. The thresholds for the single ancilla method (dotted lines) are $p_\mathrm{th}^*=0.274(6)\%$ and $p_\mathrm{th}^\times=0.2547(8)\%$. (b) Logical $Z$ error rates. The thresholds for the proposed method (solid lines) are $p_\mathrm{th}^*=0.363(9)\%$ and $p_\mathrm{th}^\times=0.352(1)\%$. The thresholds for the single ancilla method (dotted lines) are $p_\mathrm{th}^*=0.270(6)\%$ and $p_\mathrm{th}^\times=0.2528(8)\%$.}
  \label{fig:hexagonal_error_rate}
\end{figure*}
Logical $X$ and $Z$ error rates for each physical error rate are shown in Fig.~\ref{fig:square_octagonal_error_rate} for the (4.8.8) color code, and in Fig.~\ref{fig:hexagonal_error_rate} for the (6.6.6) color code.
For each, we fit the logical error rates with Eqs.~\eqref{fitting_low_error_rate} and \eqref{fitting_near_threshold} using the data for $d=7$ and $d=9$.
When we fit them using Eq.~\eqref{fitting_low_error_rate}, we use the data with a physical error rate $p$ well below the threshold, that is, in the range of $p \lessapprox p_\mathrm{th}^{\times}/2$ \cite{lee2024color}.
We use the data in the range of $p \in [10^{-4},10^{-3}]$ except for the single ancilla method for the (4.8.8) color code, where the range is $p \in [10^{-4}, 5 \times 10^{-4}]$ so that the data is considered to be well below the threshold.
When we fit the logical error rates using Eq.~\eqref{fitting_near_threshold}, we used five data points around the threshold.
The values obtained by fitting the logical $X$ and $Z$ error rates of the (4.8.8) color code in Fig.~\ref{fig:square_octagonal_error_rate} are shown in Tables~\ref{table:square_octagonal_parameter_x} and \ref{table:square_octagonal_parameter_z}, respectively.
For the (6.6.6) color code, the values obtained by fitting the logical $X$ and $Z$ error rates in Fig.~\ref{fig:hexagonal_error_rate} are shown in Tables~\ref{table:hexagonal_parameter_x} and \ref{table:hexagonal_parameter_z}, respectively.
As shown in Table~\ref{table:square_octagonal_parameter_x}, the $X$ error scaling threshold of the proposed method for the (4.8.8) color code is $p_\mathrm{th}^*$=0.29(1)\%, which is almost 1.8 times higher than that of the single ancilla method.
From Table~\ref{table:square_octagonal_parameter_z}, it can be seen that the $Z$ error scaling threshold of the proposed method for the (4.8.8) color code is $p_\mathrm{th}^*$=0.268(9)\%, which is almost 1.9 times higher than that of the single ancilla method.
The $Z$ error scaling threshold is slightly lower than the $X$ error scaling threshold, mainly because there is no flag information for the $Z$ data errors in the initial time step.
Thus, the $Z$ error scaling threshold of $p_\mathrm{th}^*$=0.268(9)\% sets the overall scaling threshold.
As shown in Tables~\ref{table:hexagonal_parameter_x} and \ref{table:hexagonal_parameter_z}, the $X$ and $Z$ error scaling thresholds of the proposed method for the (6.6.6) color code are $p_\mathrm{th}^*$=0.37(1)\% and $p_\mathrm{th}^*$=0.363(9)\%, respectively.
The $X$ error scaling threshold is almost 1.4 times higher, and the $Z$ error scaling threshold is almost 1.3 times higher than that of the single ancilla method, respectively.
In all cases, the obtained cross threshold is nearly the same as the scaling threshold.

Compared to the thresholds reported in the previous studies, for the (4.8.8) color code, we surpassed the thresholds of all previous studies \cite{landahl,graphmatching}.
For the (6.6.6) color code, we achieved the threshold that is the same within statistical errors as the highest threshold of 0.37(1)\% \cite{PRXQuantum.2.020341}, among the previous studies that adopt the same noise model.
Note that in Ref.~\cite{zhang2023facilitating}, a cross threshold of around 0.47\% was achieved, but Ref.~\cite{zhang2023facilitating} adopts a noise model that assumes fewer errors in state preparation and measurement than the noise model we adopt.
Ref.~\cite{lee2024color} obtained the cross threshold of around 0.46\%, but the obtained scaling threshold is around 0.32\%, which is lower than our value of 0.36\%.
It should also be noted that a high threshold does not necessarily guarantee lower logical error rates far below the threshold, because the logical error rates are also influenced by the effective code distance.
\begin{table}[t]
  \caption{Fitting parameters for the logical $X$ error rates of the (4.8.8) color code.}
  \label{table:square_octagonal_parameter_x}
  \hbox to\hsize{\hfil
    \begin{tabular}{ccccc}\hline\hline
                            & $c$      & $\alpha$ & $p_\mathrm{th}^*$ & $p_\mathrm{th}^\times$\\\hline
      Proposed       & 0.13(2)  &   0.69(1)     &  0.0029(1)&  0.00279(1) \\
      Single ancilla & 0.035(6) &   0.47(1)    & 0.0016(1) & 0.001694(5) \\\hline
    \end{tabular}\hfil}
\end{table}
\begin{table}[t]
  \caption{Fitting parameters for the logical $Z$ error rates of the (4.8.8) color code.}
  \label{table:square_octagonal_parameter_z}
  \hbox to\hsize{\hfil
    \begin{tabular}{ccccc}\hline\hline
                            & $c$      & $\alpha$ &  $p_\mathrm{th}^*$& $p_\mathrm{th}^\times$\\\hline
      Proposed       & 0.11(1)  &  0.679(9)     &  0.00268(9)&0.00276(1)\\
      Single ancilla & 0.031(5) & 0.49(1)      &  0.0014(1)&0.00164(2)\\\hline
    \end{tabular}\hfil}
\end{table}

In terms of the effective code distance, the proposed method improves $\alpha$ in either case.
Thereby, in both Figs.~\ref{fig:square_octagonal_error_rate} and \ref{fig:hexagonal_error_rate}, the logical error rates are almost one order of magnitude lower than those of the single ancilla method for the same code distances when the physical error rate is low, i.e., around $p=10^{-4}$.
Nonetheless, when the number of available physical qubits is given, comparing logical error rates of each method for the same code distance is not always a fair comparison.
For the range of code distances we used for the simulation, i.e., $d \in \{5,7,9\}$, the total number of data and ancilla qubits satisfies the following relationships:
\begin{equation}
    n_{4.8.8}^{\text{FWO}}(d-2) < n_{4.8.8}^{\text{single}}(d) < n_{4.8.8}^{\text{FWO}}(d),
\end{equation}
\begin{equation}
    n_{6.6.6}^{\text{FWO}}(d-2) < n_{6.6.6}^{\text{single}}(d) < n_{6.6.6}^{\text{FWO}}(d).
\end{equation}
Comparing logical error rates of each method with the same code distance $d$ is justified only when the number of available physical qubits is assumed to be $n'$ such that $n_{4.8.8}^{\text{FWO}}(d) < n' < n_{4.8.8}^{\text{single}}(d+2)$ ($n_{6.6.6}^{\text{FWO}}(d) < n' < n_{6.6.6}^{\text{single}}(d+2)$).
However, in situations where $n'$ physical qubits such that $n_{4.8.8}^{\text{single}}(d) < n' < n_{4.8.8}^{\text{FWO}}(d)$ ($n_{6.6.6}^{\text{single}}(d) < n' < n_{6.6.6}^{\text{FWO}}(d)$) are allowed to be used, the logical error rates of the single ancilla method with distance $d$ should be compared with the proposed method with distance $d-2$.
From Fig.~\ref{fig:square_octagonal_error_rate}, it can be seen that the logical error rates of the proposed method with distance 5 (7) are lower than those of the single ancilla method with distance 7 (9) for the (4.8.8) color code, which means our proposed method improves the logical error rates compared to the single ancilla method even if the number of available qubits is limited to any specific number.
In the case of the (6.6.6) color code, when the physical error rate is $p=10^{-4}$, the logical error rate of the proposed method with distance 5 is slightly higher than that of the single ancilla method with distance 7. 
However, in other physical error rates regimes, the logical error rates of the proposed method with distance 5 (7) are lower than or same as those of the single ancilla method with distance 7 (9).
\begin{table}[t]
  \caption{Fitting parameters for the logical $X$ error rates of the (6.6.6) color code.}
  \label{table:hexagonal_parameter_x}
  \hbox to\hsize{\hfil
    \begin{tabular}{ccccc}\hline\hline
                            & $c$      & $\alpha$ &  $p_\mathrm{th}^*$& $p_\mathrm{th}^\times$\\\hline
      Proposed        & 0.12(1)  & 0.72(1)      &  0.0037(1) &0.003575(6) \\
      Single ancilla & 0.116(6) & 0.610(5)      &  0.00274(6)&0.002547(8)\\\hline
    \end{tabular}\hfil}
\end{table}
\begin{table}[t]
  \caption{Fitting parameters for the logical $Z$ error rates of the (6.6.6) color code.}
  \label{table:hexagonal_parameter_z}
  \hbox to\hsize{\hfil
    \begin{tabular}{ccccc}\hline\hline
                            & $c$      & $\alpha$ &  $p_\mathrm{th}^*$& $p_\mathrm{th}^\times$\\\hline
      Proposed    & 0.117(7) & 0.675(6)      &  0.00363(9)&0.00352(1)\\
      Single ancilla & 0.111(6) & 0.600(5)      &  0.00270(6)&0.002528(8)\\\hline
    \end{tabular}\hfil}
\end{table}
\section{Conclusion}
\label{conclusion}
In this work, we have proposed flagged weight optimization (FWO), a decoder weight optimization method using conditional error probabilities conditioned on the measurement outcomes of flag qubits.
Utilizing flag values allows us to set more optimized weights, leading to more accurate decoding.
Also, the cat states reduce the depth of the syndrome measurement circuit and thus suppresses the impact of errors, as also noted in Ref.~\cite{graphmatching}.
By applying this method to the integer programming decoder, we improved the circuit-level threshold of the (4.8.8) color code from the existing 0.14\% to around 0.27\%.
In the case of the (6.6.6) color code, we achieved the circuit-level threshold of around 0.36\%, which is identical within statistical errors to the highest value of 0.37(1)\% obtained in the previous works that employ the same noise model.
In both cases, an effective code distance is also improved compared to the single ancilla method, meaning that FWO helps correct large hook errors that arise from relatively few faults. 
Thereby, the achieved logical error rates at low physical error rates are almost one order of magnitude lower than the single ancilla method with the same code distance. 
We note that the threshold values obtained here are calculated by using the code distances up to $d=9$. A numerical experiment at larger code distances is very important, but it is left for future work.

We also verified, even when comparing to the single ancilla method with a code distance higher than our method but requires the similar number of qubits, our method achieves lower logical error rates in most cases.
By utilizing this approach, it is expected that color code-based FTQC, which enables the transversal implementation of all logical Clifford gates, will become more promising.
This method can be applied to other weight-based decoders.
One can also use this method to improve the threshold of wider classes of QECCs, such as high-rate quantum LDPC codes, which have high-weight stabilizer generators.

Regarding the use of cat states, cat states offer benefits such as reducing the connectivity requirements for hardware, increasing the effective code distance, and decreasing the circuit depth. 
However, they also increase the number of qubits and circuit width.
Considering the increased number of qubits, it is generally an open question as to whether using cat states for measuring syndrome is worth it, purely from the perspective of improving logical error rates.
Nonetheless, we verified that they are worth it for improving logical error rates when performing FWO and deflagging, even taking into account the increased number of qubits, as discussed in Sec.~\ref{numerical simulation}.

In this work, we used flag information to optimize the edge weights in the decoding graph. 
On the other hand, one can use flag information as additional parity checks by adding nodes corresponding to each flag qubit and edges connecting these nodes to other relevant nodes in the decoding graph.
That method could achieve lower logical error rates than FWO, but FWO has an advantage over it. 
If we add nodes corresponding to each flag qubit and relevant edges to the decoding graph, the computational complexity of the decoding increases considerably, and more importantly, the decoding algorithm no longer works in certain decoders due to being unable to handle the additional nodes and edges. 
On the other hand, in FWO, we do not add additional nodes and edges to the decoding graph. 
Hence, if a decoder works in the phenomenological noise model, we can use FWO with that decoder in the circuit-level noise model. 
We believe that this advantage is important, because a decoder that is more promising than existing decoders in terms of decoding time or accuracy, but with limitations on the decoding graph, may be developed in the future. 
Therefore, a versatile method that is applicable to a wider variety of decoders is useful.
\begin{acknowledgments}
  The authors thank Theerapat Tansuwannont for helpful discussions.
  This work is supported by
  MEXT Quantum Leap Flagship Program (MEXT Q-LEAP)
  Grant No. JPMXS0118067394 and JPMXS0120319794, JST
  COI-NEXT Grant No. JPMJPF2014, and JST Moonshot
  R\&D Grant No. JPMJMS2061.
\end{acknowledgments}

\appendix
\section{Details of weight}
\label{appendix:a}
We describe the derivation of the decoder weight defined by Eqs.~\eqref{weight_data_previous} and \eqref{weight_meas_previous} \cite{planar}.
We consider the situation where data errors and measurement errors occur independently, with each error not having identical probabilities.
In this situation, the probability of a certain error event $E$ occurring is given by
\begin{align}
  p(E) & = \prod_{e_i \in E} p(e_i) \prod_{e_i \notin E} (1 - p(e_i))           \\
       & = \prod_{e_i \in E} \frac{p(e_i)}{1 - p(e_i)} \prod_{e_i} (1 - p(e_i)) \\
       & \propto \prod_{e_i \in E} \frac{p(e_i)}{1 - p(e_i)},
  \label{probability_of_error_event}
\end{align}
where $e_i$ denotes the $i$-th data error or measurement error.
The task of decoding is to estimate the most likely error given the syndrome $S$; that is to estimate
\begin{equation}
  E(S)=\underset{E}{\text{arg max}}\ p(E|S).
\end{equation}
Because of the monotonicity of the logarithm function, $E(S)$ can also be represented as
\begin{equation}
  E(S)=\underset{E}{\text{arg min}}\ \left( -\log p(E|S) \right).
\end{equation}
Thus, under the assumption that error probabilities are not identical across all data errors and measurement errors, the task of decoding is to estimate the errors satisfying the syndrome constraints and minimizing the sum of the following values:
\begin{equation}
  w_i=-\log \frac{p(e_i)}{1-p(e_i)}.
  \label{weight}
\end{equation}
Therefore, by setting the weights of the decoder to the values defined by Eq.~\eqref{weight}, it is possible to decode taking into account the difference in each error probability.
\setcounter{figure}{17}
\begin{figure*}[t]
  \centering
  \includegraphics[width=\linewidth]{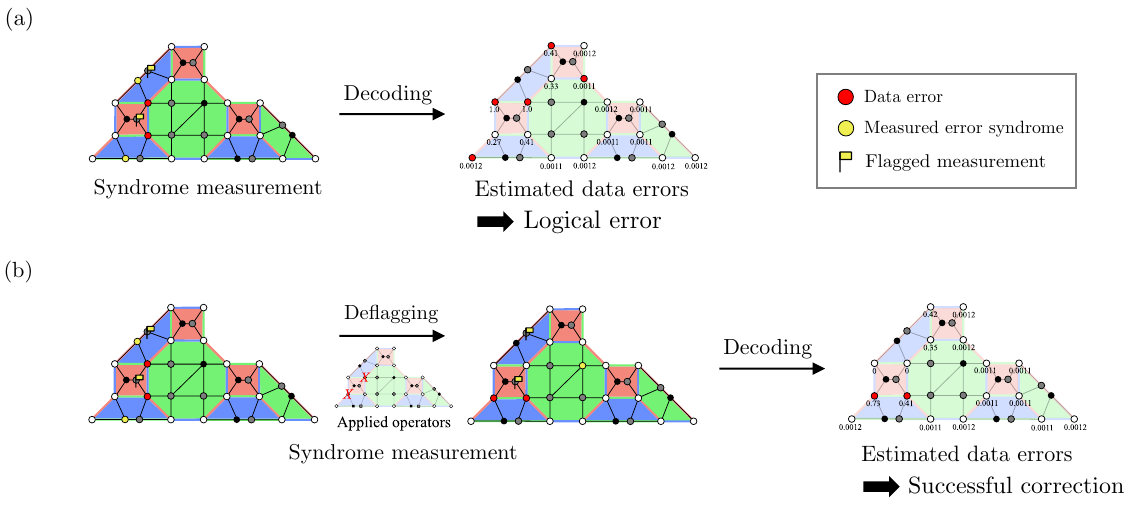}
  \caption{A typical example of performance improvement by the deflagging procedure in the case of $X$ error correction. In this figure, data errors are $X$ type, and the syndromes and flags are the ones that indicate the occurrence of $X$ errors. (a) Without deflagging. When errors occur in the syndrome measurement circuit and cause the measurement outcomes shown in the left figure, decoding based on the conditional error probabilities in Fig.~\ref{fig:deflag_estimate}(\hyperref[fig:deflag_estimate]{a}) leads to the estimation of errors as depicted in the right figure. It results in a logical error after the recovery operation. The actual data errors are also displayed in the left figure, even though they cannot be identified in actual scenarios. (b) With deflagging. After performing the deflagging procedure and then decoding based on the conditional error probabilities in Fig.~\ref{fig:deflag_estimate}(\hyperref[fig:deflag_estimate]{b}), errors are estimated as illustrated in the rightmost figure, leading to a successful error correction.}
  \label{fig:deflag_decode}
\end{figure*}
\setcounter{figure}{16}
\begin{figure}[t]
  \centering
  \includegraphics[width=\linewidth]{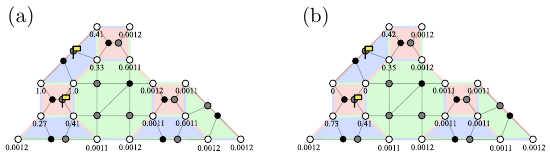}
  \caption{A typical example of how the deflagging procedure changes the conditional error probabilities. (a) Without deflagging. (b) With deflagging. The yellow flags in the figure mean the flagged measurements. The values represent the conditional error probabilities for each data error when only the two flags are triggered. The number of samplings to estimate the conditional error probabilities is $10^6$.}
  \label{fig:deflag_estimate}
\end{figure}
\setcounter{figure}{18}
\begin{figure}[tb]
  \centering
  \includegraphics[width=0.9\linewidth]{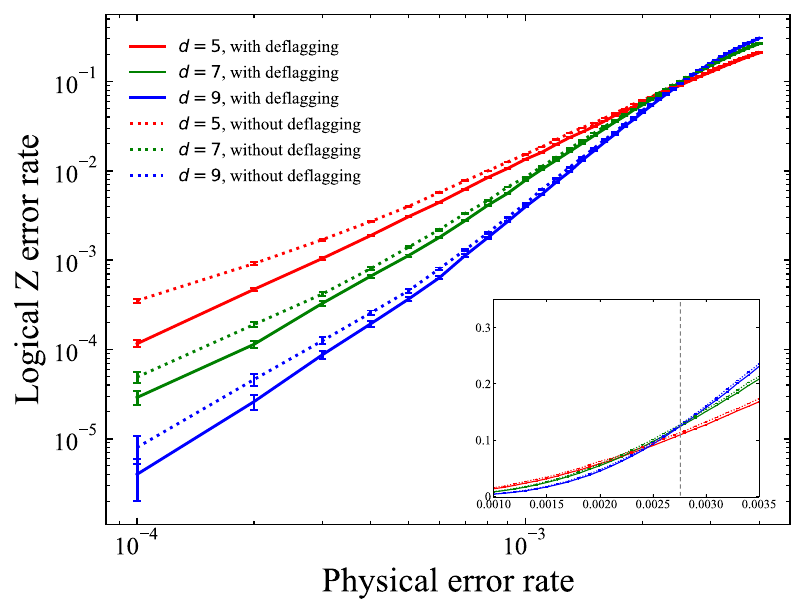}
  \caption{Logical $Z$ error rates of the (4.8.8) color code with and without deflagging procedure.}
  \label{fig:square_octagonal_error_rate_without_deflagging_z}
\end{figure}
%
\section{Details of deflagging}
\label{appendix:b}
Here, details of the deflagging procedure performed in this work are described.
In the two-qubit flag gadget, we employ a slightly different way of applying Pauli operators from Refs.~\cite{PhysRevLett.128.110504,sundaresan2023demonstrating}.
In Refs.~\cite{PhysRevLett.128.110504,sundaresan2023demonstrating}, when a flag is triggered, a Pauli operator is applied to one of the data qubits connected to the flag qubit.
In this study, when a flag is triggered in the two-qubit flag gadget, a proper type of Pauli operator is applied to all data qubits connected to the syndrome qubit.
Here, proper type means $X$ type for the flags triggered during the $X$ stabilizer measurement and $Z$ type for those triggered during the $Z$ stabilizer measurement.
When performing FWO in addition to the deflagging procedure, we verified that the improvement in logical error rates by our deflagging procedure is the same within the range of error bars as that when we perform the deflagging procedure proposed in Refs.~\cite{PhysRevLett.128.110504,sundaresan2023demonstrating}. Thus, either way of deflagging procedure can be used.
In the four-qubit flag gadget, if all three flags are triggered, a proper type of Pauli operator is applied to all data qubits connected to the top flag qubit and the syndrome qubit in Fig.~\ref{fig:flag_gadget}(\hyperref[fig:flag_gadget]{b}).

We show a typical error event that can be corrected by performing a deflagging procedure in addition to FWO.
We provide an example using the distance-5 (4.8.8) color code when the physical error rate is $p=10^{-4}$.
Let us consider a situation where errors occur in only one time step and not in other time steps to show a typical example.
Here, the weights of edges corresponding to measurement errors are not shown for simplicity.
We consider the case of $X$ error correction; that is, we focus on $X$ data errors, and the syndromes and flags considered below indicate the occurrence of $X$ errors.
The conditional data error probabilities estimated in our numerical simulation when certain two flags are triggered are shown in Fig.~\ref{fig:deflag_estimate}.
Figs.~\ref{fig:deflag_estimate}(\hyperref[fig:deflag_estimate]{a}) and \ref{fig:deflag_estimate}(\hyperref[fig:deflag_estimate]{b}) present the conditional data error probabilities without and with the deflagging procedure, respectively.
Note that estimated conditional probabilities have statistical fluctuations depending on the number of samples and the probabilities that the set of flags is triggered.
Suppose some errors occurred in the syndrome measurement circuit, resulting in the measurement outcomes shown on the left in Fig.~\ref{fig:deflag_decode}(\hyperref[fig:deflag_decode]{a}).
The actual data errors are also shown, although they are unknown in real situations.
When decoding is performed with the weights determined by the conditional error probabilities of Fig.~\ref{fig:deflag_estimate}(\hyperref[fig:deflag_estimate]{a}), an error event on the right of Fig.~\ref{fig:deflag_decode}(\hyperref[fig:deflag_decode]{a}) is estimated.
Performing a recovery operation based on this decoding result leads to a logical error.
On the other hand, when the deflagging procedure is performed for this error event, it becomes the central figure in Fig.~\ref{fig:deflag_decode}(\hyperref[fig:deflag_decode]{b}).
Note that the Pauli operators applied in the deflagging procedure here are $X$ type, because we are considering the case of $X$ error correction in this example.
Decoding with the weights determined by the conditional error probabilities in Fig.~\ref{fig:deflag_estimate}(\hyperref[fig:deflag_estimate]{b}) estimates an error event on the rightmost figure of Fig.~\ref{fig:deflag_decode}(\hyperref[fig:deflag_decode]{b}).
This leads to a successful error correction.
What we have explained above is just one typical example, so we compare the logical error rates to verify if the deflagging procedure improves the performance.
Fig.~\ref{fig:square_octagonal_error_rate_without_deflagging_z} shows the logical $Z$ error rates for the (4.8.8) color code in cases where the deflagging procedure is performed and not performed.
Fig.~\ref{fig:square_octagonal_error_rate_without_deflagging_z} indicates that the deflagging procedure improves the logical error rates.
\section{Analysis of the improvement achieved by FWO}
\label{appendix:c}
\begin{figure}[tb]
  \centering
  \includegraphics[width=0.9\linewidth]{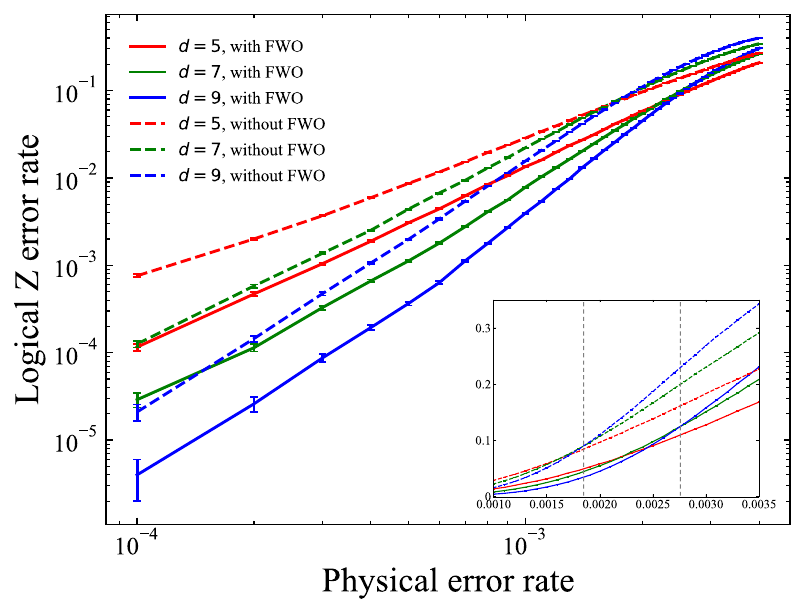}
  \caption{Logical $Z$ error rates of the (4.8.8) color code with and without FWO. Obtained scaling threshold and cross threshold for the data of the dotted lines are $p_\mathrm{th}^*=0.183(2)\%$ and $p_\mathrm{th}^\times=0.1843(9)\%$, respectively.}
  \label{fig:square_octagonal_error_rate_without_FWO_z}
\end{figure}

The numerical results in Sec.~\ref{numerical simulation} are the results of the combined contributions from several techniques described in Sec. \ref{proposed_method}, namely the use of cat states, FWO, and deflagging. 
For that reason, it is unclear how much the FWO, the main proposed technique, contributes to the results. 
Thus, we here calculate the logical error rates when using cat states and performing deflagging but without FWO, and compare with the results obtained in Sec.~\ref{numerical simulation} to clarify how much FWO itself improves the performance.
Logical $Z$ error rates for the (4.8.8) color code without performing FWO, but employing cat states and deflagging are shown in the dotted lines of Fig.~\ref{fig:square_octagonal_error_rate_without_FWO_z}.
The solid lines in Fig.~\ref{fig:square_octagonal_error_rate_without_FWO_z} and the solid lines in Fig.~\ref{fig:square_octagonal_error_rate}(\hyperref[fig:square_octagonal_error_rate]{b}) are the same data.
We also show the values obtained by fitting the logical error rates of the dotted lines in Fig.~\ref{fig:square_octagonal_error_rate_without_FWO_z} with Eqs.~\eqref{fitting_low_error_rate} and \eqref{fitting_near_threshold} in Table~\ref{table:square_octagonal_parameter_without_FWO_z}.
From Fig.~\ref{fig:square_octagonal_error_rate_without_FWO_z}, it can be seen that the logical error rates considerably decrease by performing FWO. 
Also, by comparing Tables~\ref{table:square_octagonal_parameter_without_FWO_z} and \ref{table:square_octagonal_parameter_z}, we can see that $p_\mathrm{th}^*$, $p_\mathrm{th}^{\times}$, and the effective code distance are all greatly improved by FWO. 
These results indicate that FWO itself contributes significantly to the improvements observed in Sec.~\ref{numerical simulation}.
\begin{table}[b]
  \caption{Fitting parameters for the logical $Z$ error rates of the (4.8.8) color code without FWO.}
  \label{table:square_octagonal_parameter_without_FWO_z}
  \hbox to\hsize{\hfil
    \begin{tabular}{ccccc}\hline\hline
                            & $c$      & $\alpha$ & $p_\mathrm{th}^*$ & $p_\mathrm{th}^\times$\\\hline
      Without FWO & 0.090(3) &   0.583(3)    & 0.00183(2) & 0.001843(9) \\\hline
    \end{tabular}\hfil}
\end{table}

\newpage
\bibliography{main}
\end{document}